\documentclass[12pt,amsbsy]{article}
\usepackage{amsmath}
\usepackage{amssymb}
\usepackage{graphicx}
\usepackage{psfig}

\newcommand{\beq}{\begin{equation}}
\newcommand{\eeq}{\end{equation}}
\newcommand{\beqn}{\begin{eqnarray}}
\newcommand{\eeqn}{\end{eqnarray}}

\newcommand{\lsim}{\mbox{$<$\hspace{-0.8em}\raisebox{-0.4em}{$\sim$}}}

\newcommand{\al}{\mbox{${\alpha}$}}

\newcommand{\ga}{\mbox{${\gamma}$}}
\newcommand{\Ga}{\mbox{${\Gamma}$}}

\newcommand{\De}{\mbox{${\Delta}$}}

\newcommand{\om}{\mbox{${\omega}$}}

\newcommand{\pa}{\mbox{${\partial}$}}

\begin{document}

\begin{titlepage}

\vspace{1cm}

\begin{center}
{\large \bf Quantized Black Holes,\\ Their Spectrum and Radiation}
\end{center}

\begin{center}
I.B. Khriplovich\footnote{khriplovich@inp.nsk.su}
\end{center}
\begin{center}
Budker Institute of Nuclear Physics\\
630090 Novosibirsk, Russia
\end{center}

\bigskip

\begin{abstract}
Under quite natural general assumptions, the following results are
obtained. The maximum entropy of a quantized surface is
demonstrated to be proportional to the surface area in the
classical limit. The general structure of the horizon spectrum is
found. The discrete spectrum of thermal radiation of a black hole
fits the Wien profile. The natural widths of the lines are much
smaller than the distances between them. The total intensity of
the thermal radiation is estimated.

In the special case of loop quantum gravity, the value of the
Barbero -- Immirzi parameter is found. Different values for this
parameter, obtained under additional assumption that the horizon
is described by a U(1) Chern -- Simons theory, are demonstrated to
be in conflict with the firmly established holographic bound.
\end{abstract}

\vspace{9cm}

\end{titlepage}

\section{Introduction}

The idea of quantizing the horizon area of black holes was put
forward by Bekenstein in the pioneering article~\cite{bek}. He
pointed out that reversible transformations of the horizon area of
a nonextremal black hole found by Christodoulou and
Ruffini~\cite{ch,chr} have an adiabatic nature. Of course, the
quantization of an adiabatic invariant is perfectly natural, in
accordance with the correspondence principle.

Later, the quantization of black holes was discussed by
Mukhanov~\cite{muk} and Kogan~\cite{kog}. In particular, Kogan was
the first to investigate this problem within the string approach.

Once the idea of area quantization is accepted, the general
structure of the quantization condition for large quantum numbers
$N$ gets obvious, up to an overall numerical constant (written
usually as $8\pi\,\ga$). It should be~\cite{kh}
\begin{equation}\label{qu0}
A=8\pi \,\ga \, l^2_p \, N.
\end{equation}
Indeed, the presence of the Planck length squared $l^2_p = k \hbar
/c^3$ is only natural in this quantization rule. Then, for the
horizon area $A$ to be finite in the classical limit, the power of
$N$ should be the same as that of $\hbar$ in $l^2_p$. This
argument can be checked by considering any expectation value in
quantum mechanics, nonvanishing in the classical limit.

It is worth mentioning that, contrary to widely spread beliefs,
there are no compelling reasons to believe that the black hole
spectrum (\ref{qu0}) is equidistant.

A quite popular argument in favor of its equidistance is as
follows~\cite{bem} (see also~\cite{bemu,be1}). On the one hand,
the entropy $S$ of horizon is related to its area $A$ through the
Bekenstein -- Hawking relation
\beq\label{BH}
A = 4 l_p^2 S.
\eeq
On the other hand, the entropy is nothing but $\ln g(n)$, where
the statistical weight $g(n)$ of any quantum state $n$ is an
integer. In~\cite{bem} this circumstance is used too naively,
which results after simple reasoning not only in the equidistant
spectrum (\ref{qu0}), but also in the following allowed values for
the numerical factor therein:
\[
8\pi\ga = 4 \ln k, \quad k=2,\, 3, \, ...\, .
\]
It is well-known, however, that the statistical weight being an
integer, has no consequences for the entropy of macroscopic
objects. A concrete error in arguments of~\cite{bem} is pointed
out in~\cite{kh1} (see also \cite{kh2}).

There is also an observation, usually considered as an argument in
favor of the equidistant area spectrum. It is due to Bekenstein,
who demonstrated~\cite{bek1} that quantum effects result in the
following lower bound on the change of the horizon area $\De A$
under an adiabatic process:
\beq\label{da}
(\De A)_{{\rm min}}=\xi l_p^2\,,
\eeq
where $\xi$ is a numerical factor reflecting ``the inherent
fuzziness of the uncertainty relation''.\footnote{As it should be
expected, the right-hand-side of formula (\ref{da}), corresponding
to violation of adiabatic invariance, is proportional to $\hbar$,
together with the Planck length squared $l_p^2$.} We will discuss
this bound later and demonstrate that in fact it is a strong
argument in favor of the discrete spectrum of the black hole
radiation.

\section{Structure of Quantized Area of Black Hole}

Quantization condition (\ref{qu0}) can be interpreted naturally as
follows. The whole horizon area $A$ consists of patches of typical
size $\sim l_p^2\,$. Each of them can be characterized by a
quantum number $j$, and the contribution $a$ of a patch to the
area depends on $j$, $a=a(j)$. Besides, a patch can possess a
quantum number $m$, such that $a$ is independent of it. (Of
course, both $j$ and $m$ may refer in principle not to a single
quantum number each, but to sets of them.) Then, formula
(\ref{qu0}) can be rewritten as
\begin{equation}\label{qu}
A = 8\pi\ga\, l_p^2 \sum_{jm}\,a(j)\,\nu_{jm}\,,
\eeq
where $\nu_{jm}$ is the number of patches with given $j$ and $m$.

To derive general relations for the ``occupation numbers''
$\nu_{jm}$, we will use the Bekenstein~--~Hawking relation
(\ref{BH}) and the so-called holographic bound formulated in~[13
-- 15]. According to this bound, the entropy $S$ of any spherical
nonrotating system confined inside a sphere of area $A$ is bounded
by relation
\beq\label{hb}
S \leq \frac{A}{4 l_p^2}\,,
\eeq
with the equality attained only for a system which is a black
hole.

A simple intuitive argument confirming this bound is as
follows~\cite{sus}. Let us allow the discussed system to collapse
into a black hole. During the collapse the entropy increases from
$S$ to $S_{bh}$, and the resulting horizon area $A_{bh}$ is
certainly smaller than the initial confining one $A$. Now, with
the account for the Bekenstein -- Hawking relation (\ref{BH}) for
a black hole, we arrive, through the obvious chain of
(in)equalities
\[
S \leq S_{bh} = \frac{A_{bh}}{4 l_p^2} \leq \frac{A}{4 l_p^2}\,,
\]
at the discussed bound (\ref{hb}).

It should be pointed out that at least for spherically symmetric
black holes, the holographic bound has been checked by careful
analysis of various physical situations, and therefore its
validity is firmly established.

The result (\ref{hb}) can be formulated otherwise. Among the
spherical surfaces of a given area, it is the surface of a black
hole horizon that has the largest entropy.\footnote{Even
irrespective of the holographic bound, the idea that the entropy
of a black hole is maximum, looks quite natural and was used, e.g.
in a model of quantum black hole originating from dust
collapse~\cite{vaz}.}

We will consider now the ``microcanonical'' entropy $S$ of a
quantized surface defined as the logarithm of the number of states
of this surface for a fixed area $A$ (instead of fixed energy in
common problems). Obviously, this number of states $K$ depends on
the assumption concerning the distinguishability of the patches.
So, let us discuss first of all which of {\it a priori} possible
assumptions is reasonable here from the physical point of
view~\cite{kh1} (see also~\cite{kh2}).

One possibility, which at the first glance might look quite
appealing, is that of complete indistinguishability of patches. It
means that no permutation of any patches results in new states.
Under this assumption, the total number of states created by
$\nu_j= \sum_m \nu_{jm}$ patches of a given $j$ is
\beq\label{id}
K(j) = \,\frac{(\nu_j+g(j)-1)!}{\nu_j!\; (g(j)-1)!}\,;
\eeq
here $g(j)$ is the total number of possible values of $m$ for a
given $j\,$.\footnote{Perhaps, the simplest derivation of
expression (\ref{id}) is to note that it is in fact the number of
ways of distributing $\nu_j$ identical particles into $g(j)$
boxes. Then, the line of reasoning presented in \cite{lls}, \S 54,
results in formula (\ref{id}).} Under the natural assumption
$\nu_j \gg g(j)$, the partial contributions
\[
s_j=\ln K_j = g(j)\ln \nu_j
\]
to the black hole entropy $S=\sum_j s_j$ are parametrically
smaller than the corresponding partial contributions
\[
a(j)\nu_j
\]
to $A/(4l_p^2)$, in obvious conflict with the Bekenstein --
Hawking relation (\ref{BH}). Thus, with indistinguishable patches
of the same $j$, one cannot make the entropy of a black hole
proportional to its area (see also \cite{aps}).

Let us consider now the opposite assumption, that of completely
distinguishable patches. In this case the total number of states
is
\[
K = \nu\,!\,,\quad \nu = \sum_j \nu_j =  \sum_{jm} \nu_{jm}\,,
\]
with the microcanonical entropy
\[
S=\nu \ln \nu\,.
\]
Obviously, here the maximum entropy for fixed $A \sim \sum_j
a(j)\; \nu_j$ is attained with all $a(j)$ being as small as
possible. Then, in the classical limit $\nu \gg 1$, the entropy of
a black hole grows faster than its area: $A \sim \nu$, but $S =
\nu \ln \nu \sim A\ln A$. Thus, the assumption of complete
distinguishability is in conflict with the holographic bound, and
therefore should be discarded.\footnote{There is no disagreement
between this our conclusion and that of~[18 -- 20]: what is called
complete distinguishability therein, corresponds to the last
option considered by us below, which is the only reasonable one in
our opinion.}

We go over to the third conceivable scheme, which is quite popular
(see, for instance,~\cite{aps}). According to it, the total number
of states is
\beq\label{pro}
K=\prod_j g(j)^{\nu_j},
\eeq
with the entropy of the horizon surface
\beq\label{prod}
S=\sum_j \nu_j \,\ln g(j).
\eeq
It can be easily demonstrated that this scheme corresponds to the
following assumptions on the distinguishability of patches:\\

\begin{tabular}[h]{cccc}
\vspace{3mm}
 nonequal $j$, & any $m$ & $\longrightarrow$ &
indistinguishable;\\ \vspace{3mm} equal $j$, & nonequal $m$ &
$\longrightarrow$ & distinguishable;\\
 \vspace{3mm}
equal $j$, & equal $m$ & $\longrightarrow$ & indistinguishable.\\
\end{tabular}\\

\noindent The combination of the first two of them looks strange
and unnatural (except the special case when a single value of $j$
is allowed for all patches).

Thus, we believe that the only reasonable set of assumptions on
the distinguishability of patches, which may result in acceptable
physical predictions (i.e. may comply both with the Bekenstein --
Hawking relation and with the holographic bound) is as follows:\\

\begin{tabular}[h]{cccc}
 \vspace{3mm}
 nonequal $j$, & any $m$ & $\longrightarrow$ &
distinguishable;\\ \vspace{3mm} equal $j$, & nonequal $m$ &
$\longrightarrow$ & distinguishable;\\
 \vspace{3mm}
equal $j$, & equal $m$ & $\longrightarrow$ & indistinguishable.\\
\end{tabular}\\

\noindent In this scheme, the number of states of the horizon
surface for a given number $\nu_{jm}$ of patches with momenta $j$
and their projections $j_z=m$, is obviously
\beq\label{mk}
K = \nu\,!\, \prod_{jm}\,\frac{1}{\nu_{jm}\,!}\;, \quad {\rm
where} \quad \nu =\sum_j \nu_j\,, \quad \nu_j = \sum_m \nu_{jm}\,,
\eeq
and the corresponding entropy equals
\beq\label{ms}
S=\ln K = \ln(\nu\,!)\,- \sum_{jm}\,\ln(\nu_{jm}\,!)\,.
\eeq
The structures of the last expression and of formula (\ref{qu})
are so different that in a general case the entropy certainly
cannot be proportional to the area. However, this is the case for
the maximum entropy in the classical limit.

In this limit, with all effective ``occupation numbers'' large,
$\nu_{jm} \gg 1$, we use the Stirling approximation, so that the
entropy is
\beq\label{en2}
S= \nu \ln \nu -\sum_{jm} \nu_{jm} \ln \nu_{jm}\,.
\eeq
We calculate its maximum for a fixed area $A$, i.e. for a fixed
sum\footnote{These calculations are rather obvious generalization
of those presented by us previously in~\cite{kh1,kh2,kk,kh3} for
definite structures of $a(j)$ and $g(j)$ predicted in loop quantum
gravity (see section 5).}
\beq\label{N}
N\,=  \sum_{jm}^\infty a(j)\,\nu_{jm}={\rm const} \,.
\eeq

The problem reduces to the solution of the system of equations
\beq\label{sys}
\ln \nu  - \ln \nu_{jm} = \mu\, a(j)\,,
\eeq
where $\mu$ is the Lagrange multiplier for the constraining
relation (\ref{N}). These equations can be rewritten as
\beq\label{nu1}
\nu_{jm}=\nu\, e^{- \mu\, a(j)},
\eeq
or
\beq\label{nu2}
\nu_j = \,\nu\,g(j)\,e^{- \mu\, a(j)}.
\eeq
Now we sum expressions (\ref{nu2}) over $j$, and with $\sum_j
\nu_j = \nu$ arrive at the equation for $\mu$:
\beq\label{equ}
\sum_j g(j)\, e^{- \mu \,a(j)} = 1.
\eeq

Strictly speaking, the summation in formula (\ref{equ}) extends
not to infinity, but to some $j$ corresponding to the maximum
contribution $a_{{\rm max}}$ to the area. The value of $a_{{\rm
max}}$ follows from the obvious condition: none of $\nu_{jm}$
should be less than unity. Then, for $\nu \gg 1$ equation
(\ref{nu1}) gives
\beq\label{jm}
a_{{\rm max}} = \,\frac{\ln \nu}{\mu}\,.
\eeq
It is well-known that the Stirling approximation for $n!$ has
reasonably good numerical accuracy  even for $n = 1$. Due to it,
formula (\ref{jm}) for $a_{{\rm max}}$ is not just an estimate,
but has reasonably good numerical accuracy.

On the other hand, multiplying equation (\ref{sys}) by $\nu_{jm}$
and summing over $jm$, we arrive, with the constraint (\ref{N}),
at the following result for the maximum entropy for given $N$:
\beq\label{enf0}
S_{\rm max}= \,\mu \,N\,=\,\frac{\mu}{8\pi\ga l_p^2}\,A.
\eeq

Thus, equation (\ref{qu}) for the quantized area can be rewritten
as
\beq\label{qu1}
A = 8\pi\ga\, l_p^2\, \nu \sum_j\,g(j)\,a(j)\, e^{- \mu a(j)}\,,
\eeq
where $\ga = \mu/(2\pi)$, and the value of $\mu$ is found from
equation~(\ref{equ}).

\section{Quantization of Mass of Rotating Black Hole}

When discussing the radiation spectrum of quantized black holes,
one should take into account the selection rules for angular
momentum. Obviously, radiation of any particle with nonvanishing
spin is impossible if both initial and final states of a black
hole are spherically symmetric. Therefore, to find the radiation
spectrum, the quantization rule for the mass of a Schwarzschild
black hole should be generalized to that of a rotating Kerr black
hole.

To derive the quantization rule for Kerr black hole, we come back
to the thought experiment analyzed in~\cite{ch,chr}. Therein,
under the adiabatic capture of a particle with an angular momentum
$j$, the angular momentum $J$ of a rotating black hole changes by
a finite amount $j$, but the horizon area $A$ does not change. Of
course, under some other variation of parameters it is the angular
momentum $J$ that remains constant. In other words, we have here
two independent adiabatic invariants, $A$ and $J$, for a Kerr
black hole with a mass $M$.

Such a situation is quite common in ordinary mechanics. For
instance, the energy of a particle with mass $m$ bound in the
Coulomb field $U(r) = - \al/r$ is
\beq\label{Ec}
E=-\,\frac{m \al^2}{2\,(I_r + I_{\phi})^2}\,,
\eeq
where $I_r$ and $I_{\phi}$ are adiabatic invariants for the radial
and angular degree of freedom, respectively. Of course, the energy
$E$ is in a sense an adiabatic invariant also, but it is invariant
only with respect to those variations of parameters under which
both $I_r$ and $I_{\phi}$ remain constant (or at least their sum).
As to quantum mechanics, in it formula (\ref{Ec}) goes over into
\beq\label{Eq}
E=-\,\frac{m \al^2}{2\,\hbar^2\, (n_r + 1 + l)^2}\,,
\eeq
where $n_r$ and $l$ are the radial and orbital quantum numbers,
respectively.

This example prompts the solution of the quantization problem for
a Kerr black hole. It is conveniently formulated in terms of the
so-called irreducible mass $M_{ir}$ of a black hole, related by
definition to its horizon radius $r_h$ and area $A$ as follows:
\beq\label{rel}
r_h= 2 k M_{ir}\,, \quad A = 16\pi k^2 M_{ir}^2\,.
\eeq
Together with the horizon area $A$, the irreducible mass is an
adiabatic invariant. In accordance with (\ref{qu}) and (\ref{N}),
it is quantized as follows:
\beq\label{mirq}
M_{ir}^2 =\,\frac{1}{2}\,m^2_p \,N\,,
\eeq
where $m^2_p=\hbar c/k$ is the Planck mass squared.

Of course, for a Schwarzschild black hole $M_{ir}$ coincides with
its ordinary mass $M$. However, for a Kerr black hole the
situation is more interesting. Here
\beq\label{mik}
M^2=M_{ir}^2 + \,\frac{J^2}{r_h^2}\,=M_{ir}^2 + \,\frac{J^2}{4k^2
M_{ir}^2}\,,
\eeq
where $J$ is the internal angular momentum of a rotating black
hole.

Now, with the account of equation (\ref{mirq}), we arrive at the
following quantization rule for the mass squared $M^2$ of a
rotating black hole:
\beq\label{mqj}
M^2 =\,\frac{1}{2}\,m^2_p \left[\ga N +\,\frac{J(J+1)}{\ga
N}\right].
\eeq
Obviously, as long as a black hole is far away from an extremal
one, i.e. while $\ga N \gg J$, one can neglect the dependence of
$M^2$ on $J$, and the angular momentum selection rules have
practically no influence on the radiation spectrum of a black
hole.

As to the mass and irreducible mass of a charged black hole, they
are related as follows:
\beq\label{mir}
M=M_{ir} + \,\frac{q^2}{2r_h}\,;
\eeq
here $q$ is the charge of the black hole. This formula has a
simple physical interpretation: the total mass (or total energy)
$M$ of a charged black hole consists of its irreducible mass
$M_{ir}$ and of the energy $q^2/2r_h$ of its electric field in the
outer space $r>r_h$.

With $r_h= 2 k M_{ir}$, relation (\ref{mir}) can be rewritten as
\beq\label{mic}
M^2=M_{ir}^2 + \,\frac{q^4}{16 k^2 M_{ir}^2} + \frac{q^2}{2k}\,.
\eeq
Thus, for a charged black hole $M^2$ is quantized as follows:
\beq\label{mqq}
M^2=\,\frac{1}{2}\,m^2_p \left[\ga N +\,\frac{q^4}{4\ga N}+ q^2
\right].
\eeq

In fact, relations of this type (even in a more general form, for
Kerr -- Newman black holes, both charged and rotating) were
presented already in the pioneering article~\cite{bek}, though
with the equidistant quantization rule for $M_{ir}^2$, i.e. for
the horizon area (see also~\cite{bek1}). More recently, the
conclusion that the mass of a quantized black hole should be
expressed via its quantized area and angular momentum, was made in
the approach based on the notion of so-called isolated
horizons~\cite{af,abl}.

I do not mention here those attempts to quantize rotating and
charged black holes which resulted in weird quantization rules for
$\hat{J}^2$ and $e^2/\hbar c$.

\section{Radiation Spectrum of Quantized Black Hole}

It follows from expression (\ref{mqj}) that for a rotating black
hole the radiation frequency $\om$, which coincides with the loss
$\De M$ of the black hole mass, is
\beq\label{om}
\om = \De M = T \mu \,\De N +\,\frac{1}{4kM}\,\frac{2J+1}{\ga N}\,
\De J\,,
\eeq
where $\De N$ and $\De J$ are the losses of the area quantum
number $N$ and of the angular momentum $J$, respectively. We have
used here, in line with (\ref{mqj}), the following identity for
the Hawking temperature $T$:
\beq
T=\,\frac{\pa M}{\pa S}\,=\,\frac{1}{8\pi k M}\,\frac{\pa M^2}{\pa
M_{ir}^2}\,,
\eeq
as well as formula (\ref{mik}).

In the same way, for a charged black hole one obtains with formula
(\ref{mqq}) the radiation frequency
\beq\label{omc}
\om = \De M = T \mu \,\De N +\,\frac{1}{4 k M}\,\left(2 +
\frac{q^2}{\ga N}\right)q\,\De q\,,
\eeq
where $\De q$ is the loss of the charge.

We will be interested mainly in the first, temperature terms in
(\ref{om}) and (\ref{omc}), dominating everywhere but the vicinity
of the extremal regime, where $J \to \ga N$, or $q^2 \to 2 \ga N$,
and $T \to 0$. The natural assumption is that the temperature
radiation occurs when a patch with a given value of $j$
disappears, which means that
\beq\label{dom}
\De N_j = a(j)\,, \quad \om_j = T \mu\, a(j)\,.
\eeq
Thus we arrive at the discrete spectrum with a finite number of
lines. Their frequencies start at $\om_{{\rm min}}=T\mu \, a_{{\rm
min}}$, where $a_{{\rm min}}$ is the minimum value of $a(j)$, and
terminate at $\om_{{\rm max}}=T\ln\nu$ (we recall here that
$a_{{\rm max}} =\ln \nu/\mu$). Thus, the number of lines is not so
large, $\sim 10^2$, if the mass of black hole is comparable to
that of the Sun. However, due to the exponential decrease of the
radiation intensity with $\om$ (see below), the existence of
$\om_{{\rm max}}$ and finite number of lines are not of much
importance.

To substantiate the made assumption, we come back to the lower
bound (\ref{da}) on the change of the horizon area under an
adiabatic capture of a particle. The presence of the gap
(\ref{da}) in this process means that this threshold capture
effectively consists in the increase by unity of the occupation
number $\nu_{jm}$ corresponding to $a_{{\rm min}}$. If the capture
were accompanied by a reshuffle of few occupation numbers, the
change of the area could be made in general as small as one
wishes.\footnote{Except the case when $a(j)$ is a linear function
of $j$, and, correspondingly, the area spectrum is equidistant,
which generally speaking cannot be excluded.}

It is only natural to assume that in the radiation process as
well, changing few occupation numbers, instead of one, is at least
strongly suppressed. In this way we arrive at
equations~(\ref{dom}).

Our next assumption, at least as natural as this one, is that the
probability of radiation of a quantum with frequency $\om_j$ is
proportional to the occupation number $\nu_j$. Correspondingly,
the radiation intensity $I_j$ at this frequency $\om_j$ is
proportional to $\nu_j\, \om_j$:
\beq\label{ij}
I_j \sim \nu_j\, \om_j \sim \nu\, g(j)\, \om_j \, e^{-\omega_j/T}.
\eeq

We compare now this expression with the intensity of the
black-body radiation in the Wien limit $\om/T \gg 1$,
\beq\label{w}
I(\om)\,d\om = A\,\frac{\om^3}{4 \pi^2}\,e^{-\omega/T}d\om,
\eeq
where $A$ is the area of a spherical black body. First of all, our
relation (\ref{ij}) for $I_j$ reproduces naturally the exponential
factor of the Wien spectrum. To reproduce the Wien profile
completely, we have to supplement relation (\ref{ij}) with an
obvious additional factor, a sort of ``oscillator strength''. Thus
we arrive at the final expression for the discrete radiation
spectrum of a black hole:
\beq\label{ija}
I_j\,\De \om_j = I_j\,\frac{\pa a(j)}{\pa j}\,\De j =
A\,T^4\,\frac{1}{4\pi^2}\,\mu^4 a^3(j)\,\frac{\pa a(j)}{\pa
j}\,e^{-\mu\,a(j)}\,\De j\,.
\eeq
Here $\De j$ is the difference between two successive values of
$j$. For instance, if only integer values of $j$ are admissible,
$\De j=1$; if half-integer values are possible as well, $\De
j=1/2\,$.

One should mention that it was argued long ago~\cite{bem}, for the
case of equidistant horizon quantization, that the discrete
thermal radiation spectrum of a black hole should fit the Wien
profile.

Of course, the discrete spectrum (\ref{ija}), together with the
Wien spectrum (\ref{w}), refers, strictly speaking, to high
frequencies $\;\om_j/T \,=\,\mu\,a(j) \gg 1$. However, one may
hope that even for $\,\mu\,a_{{\rm min}}\,\sim\,1\,$ the
asymptotic value given by (\ref{ija}) does not differ too much
from the true one.

On the other hand, when calculating here the radiation intensity,
we have, in principle, to introduce the so-called grey factor,
which is absent of course in the thermal radiation of a common
black body. However, under the natural assumption $\,\om_{\rm
min}/T\,=\,\mu\,a_{{\rm min}}\, \sim\,1\,$, the grey factor
correction should not be as essential quantitatively even for the
first line, and so much the more for higher ones.

Now, to estimate the total radiation intensity $I= \sum_j I_j\,\De
\om_j\,,$ we approximate the sum over $j$ by integral and thus
obtain
\beq\label{tot}
I \simeq A\,T^4\,\frac{1}{4\pi^2}\,\int_{\mu a_{{\rm min}}}^\infty
dx \,x^3 e^{-x}\,.
\eeq
The integral obtained is almost independent of its lower limit
$\mu a_{{\rm min}}$ if this limit remains on the order of unity:
the value of the integral changes from 6 to 5.14 when $\mu a_{{\rm
min}}$ changes from 0 to 2. Therefore, under the same natural
assumption $\mu a_{{\rm min}} \sim 1$ the total radiation
intensity is
\beq
I \approx 0.14\, A \,T^4\,.
\eeq
The numerical coefficient in this expression is close to that in
the total intensity of the common thermal radiation, i.e. to the
Stefan-Boltzmann constant, which is $\pi^2/60 =0.164$. This is
natural since the Rayleigh-Jeans contribution to the total
intensity, which is absent in the present spectrum, would be small
anyway.

The emission probability for a quantum of frequency $\om_j = T
\mu\, a(j)$, i.e. the width of the corresponding line, is
\beq\label{lw}
\Ga_j\,\De \om_j = \,\frac{I_j\,\De \om_j}{\om_j}\,=
A\,T^3\,\frac{1}{4\pi^2}\,\mu^3 a^2(j)\,\frac{\pa a(j)}{\pa
j}\,e^{-\mu\,a(j)}\,\,\De j.
\eeq
The ratio of this natural line width to the distance $\De \om_j$
is
\beq\label{rat}
\Ga_j\, =\, \frac{1}{16\pi^3}\,\left[\mu\,
a(j)\right]^2\,e^{-\mu\, a(j)}.
\eeq
With
\[
\left[\mu\, a(j)\right]^2\,e^{-\mu\, a(j)}\,\lsim \; 0.54\,,
\]
this ratio is very small numerically: $\lsim \; 10^{-3}$. Thus,
the radiation spectrum of an isolated black hole is really
discrete.

Equation (\ref{ija}) describes the thermal radiation of photons
and gravitons, which have two polarizations. It applies as well to
the thermal radiation of massless fermions. However, in the last
case a proper account of the number of polarization states is
necessary: e.g. for a two-component Dirac neutrino, with a single
polarization, the numerical factor in (\ref{ija}) will be two
times smaller.

As to the nonthermal radiation of extremal black holes, described
by the terms with $\De J$ and $\De q$ in (\ref{om}) and
(\ref{omc}), these effects are due to tunneling (see relatively
recent discussion of the subject, as well as detailed list of
relevant references, in~\cite{khr,kk1}). The loss of a charge by a
charged black hole is caused in fact by the Coulomb repulsion
between the black hole and emitted particles with the same sign of
charge. For a rotating black hole the cause is the interaction of
angular momenta: particles (massless mainly), whose total angular
momentum is parallel to that of a black hole, are repelled from
it.

\section{Black Holes in Loop Quantum Gravity}

In this section we illustrate the general relations derived above
with an example of a concrete model, that of loop quantum gravity
(LQG)~[27 -- 31].

A quantized surface in LQG looks as follows. One ascribes to it a
set of punctures (corresponding to patches of the previous
sections). Each puncture is supplied with an integer or
half-integer quantum number $j$:
\beq\label{j}
j= 1/2, 1, 3/2, ...\; .
\eeq
The projections $m$ of these ``angular momenta'' run as usual from
$-j$ to $j$. The area of a surface is
\beq\label{Aj1}
A =8\pi\ga\, l_p^2 \sum_{jm} \sqrt{j(j+1)}\,\nu_{jm}.
\eeq
This is in fact a special case of the above general expressions
with
\beq\label{ag}
a(j)\, =\, \sqrt{j(j+1)}\,, \quad g(j) = 2j+1.
\eeq

It is worth mentioning here that though area spectrum (\ref{Aj1})
is not equidistant, it is not far away from it. Indeed, even for
the smallest quantum number $j=1/2$, $\sqrt{j(j+1)}$ can be
approximated by $j+1/2$ with an accuracy 13\%. And the
approximation $\sqrt{j(j+1)} \approx j+1/2$ gets better and better
with growing $j$, i.e. spectrum (\ref{Aj1}) approaches an
equidistant one more and more.

The numerical factor $\ga$ in (\ref{Aj1}) (called here Barbero --
Immirzi parameter)  corresponds in LQG to a family of inequivalent
quantum theories, all of them being viable without such an
input~\cite{imm,rot}.

General relations derived above for arbitrary $a(j)$, $g(j)$ can
be readily used in the present concrete case. In particular,
``secular'' equation (\ref{equ}) and its solution read
now\footnote{These results were obtained by us few years
ago~\cite{kk} (see also~\cite{kh1,kh2,kh3}), and reproduced
recently in~\cite{gm}, in somewhat different context.}
\beq\label{equ1}
\sum_{j=1/2}^{\infty} (2j+1)\, e^{- \mu \sqrt{j(j+1)}} = 1\,,\quad
\mu = 1.722.
\eeq
Thus, the value of the Barbero-Immirzi parameter in LQG is
\beq\label{bip1}
\ga = \,\frac{\mu}{2\pi}\,= 0.274.
\eeq
The minimum frequency is here
\beq
\om_{{\rm min}} = \mu\,\frac{\sqrt{3}}{2}\,T = 1.491\, T\,,
\eeq

One can find also in closed form corrections to the Bekenstein --
Hawking relation (\ref{BH}). With the leading correction included,
this relation looks as follows:
\beq\label{BH1}
S
=\,\frac{A}{4l^2_p}\,-\,\frac{1}{6\mu^2}\,\ln^3\frac{A}{l^2_p}\,.
\eeq
The existence of this $\ln^3$ correction was first pointed out
in~\cite{gm}, though with coefficient 1/3 instead of 1/6.

At last, the total radiation intensity of a black hole in LQG is
\beq\label{itot}
I = \sum_j I_j \De \om_j =  A \,T^4 \,\frac{1}{8\pi^2}\,\mu^4
\sum_{j=1/2}^\infty [j(j+1)]^{3/2}\,e^{-\mu \sqrt{j(j+1)}}\,=
0.144 \,A \,T^4.
\eeq

We note that our conclusion of the discrete radiation spectrum of
a black hole in LQG differs drastically from that of~\cite{bcr},
according to which this spectrum is dense.

Our last remark refers to the problem of Barbero -- Immirzi
parameter $\ga$ in LQG. The first attempts to fix its value, based
on the analysis of the black hole entropy, were made
in~\cite{rov,kra}. However, these attempts did not lead to
concrete quantitative results.

Then it was argued in~\cite{asht} that the horizon of a black hole
should be described by a U(1) Chern -- Simons theory, and
characterized by punctures with quantum numbers $\pm 1/2$
only.\footnote{This scheme coincides effectively with the
so-called ``it from bit'' model formulated earlier by
Wheeler~\cite{whe}.} With these quantum numbers, one arrives
easily at the equidistant area spectrum and at the value $\ga_1 =
\ln 2/(\pi \sqrt{3}) = 0.127$ for the Barbero -- Immirzi (BI)
parameter.

However, it was pointed out in~\cite{kh1} that this result
violates the holographic bound. Indeed, common bulk states, where
a surface area is given by expression (\ref{Aj1}), certainly exist
in LQG. With smaller BI parameter, $\ga_1 = 0.127\, < \ga =
0.274$, the entropy obtained in~\cite{asht} is smaller than the
maximum entropy corresponding to area (\ref{Aj1}). Therefore, in
virtue of the firmly established (at least for spherically
symmetric surfaces) holographic bound, the result of~\cite{asht}
cannot be correct.

Then, the result of~\cite{asht} was revised in~\cite{lew,mei},
still within the same idea of a U(1) Chern~--~Simons description
of the horizon. Effectively, according to~\cite{lew,mei}, one
should ascribe to the punctures of the horizon arbitrary integer
and half-integer $j$, but with only two maximum projections $\pm
j$. No wonder that the equation for the BI parameter looks
in~\cite{mei} as
\beq\label{2}
2\sum_{j=1/2}^{\infty}\, e^{- \mu \sqrt{j(j+1)}} = 1
\eeq
with the solution $\ga_2 = \mu_2/(2\pi) = 0.238$, instead of ours
(\ref{equ1}) with $\ga = 0.274$ (see also the discussion of
(\ref{2}) in~\cite{gm}). So, since $ \ga_2 = 0.238 \,< \ga =
0.274$, here as well the holographic bound is violated.

The conclusion is obvious. Any restriction on the number of
admissible states for the horizon in LQG, as compared to a generic
quantized surface (with any $\,j\,$ and $\, -j\,\leq \,m \,\leq
\,j \,$), be it, for instance, the restriction to
\[
j=1/2\,, \quad m=\pm 1/2\,,
\]
according to~\cite{asht}, or the restriction to
\[
{\rm any} j\,, \quad m=\pm j\,,
\]
according to~\cite{lew,mei}, results in a conflict with the
holographic bound. Clearly, the schemes leading to such
restrictions should be abandoned.

\section*{Acknowledgements}
I am grateful to A.A. Pomeransky and I.V. Volovich for useful
discussions. The investigation was supported in part by the
Russian Foundation for Basic Research through Grant No.
05-02-16627.

\end{document}